\documentclass[letterpaper,11pt]{article}
\usepackage{osajnl2}
\usepackage[utf8]{inputenc}
\usepackage{epsfig}
\usepackage{amsmath}
\usepackage{amssymb}
\usepackage[centerlast,rm,footnotesize]{subfigure}
\begin{document}
%%%%%%%%%%%%%%%%%%%%%%%%%%%%%%%%%%%%%%%%%%
\pagestyle{empty}

%%%%%%%%%%%%%%%%%%%%%%%%%%%%%%%%%%%%%%%%%%
\title{Wake effects characterization using wake oscillator model \\ Comparison on 
2D response with experiments}
%%%%%%%%%%%%%%%%%%%%%%%%%%%%%%%%%%%%%%%%%%

%%%%%%%%%%%%%%%%%%%%%%%%%%%%%%%%%%%%%%%%%%
\author{Benoît Gaurier$^{1,*}$, Grégory Germain$^1$ and David Cébron$^{2}$}
\address{$^1$Ifremer - Bassin d'Essais, 150 quai Gambetta - BP 699 \\ 62 321 Boulogne-sur-Mer, France}
\address{$^2$Ecole Centrale Nantes, 1 rue de la Noë - BP 92101 \\ 44 321 Nantes Cedex 3, France}
\address{$^*$Corresponding author: benoit.gaurier@ifremer.fr}
%%%%%%%%%%%%%%%%%%%%%%%%%%%%%%%%%%%%%%%%%%

%%%%%%%%%%%%%%%%%%%%%%%%%%%%%%%%%%%%%%%%%%
\begin{abstract}
A model using wake oscillators is developed to predict the 2D motion in a transverse plan of two rigid cylinders in tandem arrangement. This model of the wake dynamics is validated with experimental data from previous trials which took place at the Ifremer flume tank in Boulogne-sur-Mer, France. The agreement between the model and the experimental results allows using this model as a simple computational tool in the prediction of 2D Vortex-Induced Vibrations (VIV) and, after some futher developments, Wake-Induced Oscillations (WIO) effects.
\end{abstract}
%%%%%%%%%%%%%%%%%%%%%%%%%%%%%%%%%%%%%%%%%%

\maketitle

%%%%%%%%%%%%%%%%%%%%%%%%%%%%%%%%%%%%%%%%%%
\section{Introduction}
Mooring and flow lines involved in offshore systems for oil production are submitted to various solicitations. Among them the effects of current are dominating. Vortex-Induced Vibrations (VIV) and Wake-Induced Oscillations (WIO) on closely spaced marine risers may lead to fatigue, clashes and structural failures. Extended studies have been conduced to describe and explain them for spring mounted uniform cylinders in translation perpendicularly to their main axis \cite{Sarpkaya2004}. In the case of a pivoted cylinder with uniform diameter \cite{Flemming2005} a similar response is observed with some variations on the reduced velocity interval and maximum response. 

Experiments on model scaled tests with real configurations for dual risers interaction in a uniform current were performed in the Ifremer flume tank, within the framework of the project Clarom CEPM CO 3007/04 in partnership with Doris engineering, Saipem S.A., Institut Français du Pétrole, Océanide, Ecole Centrale Marseille and Total. The behaviour of two risers exposed to steady current and activated by VIV and WIO were studied \cite{Germain2006}. These tests give a lot of information on how fluid interaction between two cylinders of equal diameter in tandem configuration can significantly modify their structural response in term of amplitude and frequency, compared to a single one \cite{Gaurier2007b}. Both in-line and cross-flow response have been studied and presented as functions of the reduced velocity. Results demonstrate that wake effects can be relatively strong. In almost all the tested cases the upstream cylinder responds like an isolated single one, whereas the vortex shedding and synchronization of the downstream cylinder can be strongly affected by the wake of the upstream one. Those phenomenons are relatively hard to predict. 

In order to quantify those wake effects, we developed a 2D phenomenological model of the near wake based on Van Der Pol wake oscillator (see \cite{Facchinetti2004} and \cite{Violette2007}) which describes the 2D motion of the cylinder in its transverse plan. This simplified model of the wake dynamics was first validated on a single cylinder in \cite{Cebron2008} and is extended here for the case of two cylinders in interaction.

%%%%%%%%%%%%%%%%%%%%%%%%%%%%%%%%%%%%%%%%%%
\section{Mathematical model}
%%%%%%%%%%%%%%%%%%%%%%%%%%%%%%%%%%%%%%%%%%
\subsection{Dynamic equations in 2D}

The equations of this mathematical model are presented in \cite{Cebron2008}. Here are just the main formulas needed for the basic comprehension.

\noindent Initially, dynamic equations are written in a generalised way for a cylinder in free motion in its transverse plan. For an oscillating cylinder (mass $m$, volume $V$, velocity $\dot{\overrightarrow{X}}$) submit to a current described by: $\overrightarrow{U}=U_x (t) \; \overrightarrow{e_x} + U_y (t) \; \overrightarrow{e_y}$, the center of inertia theorem gives:

\begin{equation}
m \; \overrightarrow{a_r} = \sum \overrightarrow{F_{ext}} + \overrightarrow{f_{ie}} + \overrightarrow{f_{ic}}
\end{equation} 

\noindent with $\overrightarrow{a_r}=\ddot{\overrightarrow{X}}-\dot{\overrightarrow{U}}$ the relative acceleration, $\overrightarrow{f_{ie}} = -m \; \dot{\overrightarrow{U}}$ the inertial force and $\overrightarrow{f_{ic}} = \overrightarrow{0}$ the Coriolis force. The exterior forces $\sum \overrightarrow{F_{ext}}$ contain the hydrodynamic forces (drag, lift and forces issued from the potential theory), the spring force and possibly structural damping forces:

\begin{equation}
\sum \overrightarrow{F_{ext}} = \overrightarrow{F_{hydro}} + \overrightarrow{f_{spring}} + \overrightarrow{f_{damping}}
\end{equation}

\bigskip
Those previous forces are those in laminar flow, without vortex. But in turbulent flow, other special forces have to be added: the fluctuating drag and lift forces created by vortex and the blockage drag which is the additional drag issued from the transverse motion of the cylinder: this motion increases the apparent projected surface in front of the flow.

\bigskip
Finally, in projection with $\overrightarrow{F_{L}} = \overrightarrow{e_{z}} \times \overrightarrow{F_{D}}$, the final general equations are:
\begin{equation} \label{struct_oscil} 
 \left\{
  \begin{array}{l@{\;}l}
    (m+C_m \rho V) \ddot{x} + \lambda \dot{x} + k(x-x_0) = \displaystyle \frac{1}{2} \rho S \left[ (C_D + C_{Df}) (U_x-\dot{x}) - (C_L +C_{Lf}) (U_y-\dot{y}) \right] \times \\ \sqrt{(U_x-\dot{x})^2+(U_y-\dot{y})^2} \displaystyle + \rho V (1+C_m) \dot{U_x} + \frac{1}{2}\rho S C_{Db}\sqrt{U_x^2+U_y^2} \; U_x \\

    \\

    (m+C_m \rho V) \ddot{y} + \lambda \dot{y} + k(y-y_0) = \displaystyle \frac{1}{2} \rho S \left[ (C_D + C_{Df}) (U_y-\dot{y}) + (C_L + C_{Lf}) (U_x-\dot{x}) \right] \times \\ \sqrt{(U_x-\dot{x})^2+(U_y-\dot{y})^2} \displaystyle + \rho V (1+C_m) \dot{U_y} + \frac{1}{2}\rho S C_{Db}\sqrt{U_x^2+U_y^2} \; U_y \\
  \end{array}
\right.
\end{equation}

\noindent with $\rho$ the mass density, $C_m$ the added mass coefficient, $C_D$ and $C_{Df}$ the average and fluctuating drag coefficient, $C_{Db}$ the blockage drag coefficient, $C_L$ and $C_{Lf}$ the average and fluctuating lift coefficient, $k$ the stiffness and $\lambda$ the linear structural damping.

%%%%%%%%%%%%%%%%%%%%%%%%%%%%%%%%%%%%%%%%%%
\subsection{Vortex forces}
%%%%%%%%%%%%%%%%%%%%%%%%%%%%%%%%%%%%%%%%%%

Following \cite{Facchinetti2004}, the vortex forces could be modeled on Van Der Pol oscillators coupled with the acceleration of the cylinder:

\begin{equation} \label{fluide_oscil} 
 \left\{
  \begin{array}{l@{\;}l}
    \displaystyle \ddot{C}_{Df} + \varepsilon_{D} \; 2 \omega_{st} \left( \left(\frac{2 \; C_{Df}}{C_{Df_0}}\right)^2-1 \right) \dot{C}_{Df} + \; (2 \omega_{C_L})^2 \; C_{Df} = A_D (\ddot{x}-\dot{U}_x) \\

    \displaystyle \ddot{C}_{Lf} + \varepsilon_{L} \; \omega_{st} \left( \left(\frac{2 \; C_{Lf}}{C_{Lf_0}}\right)^2-1 \right) \dot{C}_{Lf} + \; \omega_{C_L}^2 \; C_{Lf} = A_L (\ddot{y}-\dot{U}_y) \\
  \end{array}
\right.
\end{equation} 

\noindent where $C_{Df_0}$ and $C_{Lf_0}$ are the amplitudes of the fluctuating drag and lift coefficients, $\omega_{st}$ is the vortex shedding pulsation, issued from the Strouhal number. \cite{Facchinetti2004} and \cite{Violette2007} use $A_L=12$ and $\varepsilon_L = 0.3$ for one cylinder in 1D motion, while we use here: $A_L=7$, $A_D=1$, $\varepsilon_D = 1.2$ and $\varepsilon_L = 2.5$ for both cylinders. These parameters are issued from a preliminary study based on an optimization algorithm: the method of gradient descent.

%%%%%%%%%%%%%%%%%%%%%%%%%%%%%%%%%%%%%%%%%%
\subsection{Hydrodynamical parameters} \label{hydro}
%%%%%%%%%%%%%%%%%%%%%%%%%%%%%%%%%%%%%%%%%%

For the accuracy of the model, the evolution of hydrodynamical coefficients must be known precisely. So, bibliographical data from \cite{Sarpkaya1978}, \cite{Sarpkaya2004}, \cite{Schlichting1987}, \cite{Norberg2001} and \cite{Cantwell1983} is fitted to obtain analytic formulas, versus Reynolds number $Re=U D / \nu$ or reduced velocity $V_r = U / (f_n D)$, with $f_n$ the natural frequency of the system. All these formulas are presented in \cite{Cebron2008} and concern the mean drag coefficient $C_D$, the Strouhal number $St$, the added mass coefficient $C_m$, the correlation length $\Lambda_L$ and the fluctuating lift coefficient $C_{Lf}$.

\bigskip
Indeed, these formulas refer to only one cylinder. For the downstream cylinder, there are some differences to take into account, linked to the presence of the upstream one and of its wake. For example, on the contrary to the upstream cylinder, the average lift force of the downstream cylinder $C_{L_2}$ is not always equal to zero when this cylinder is not in line with the upstream one.

\noindent In the following lines, indice $1$ stands for the upstream cylinder, whereas indice $2$ stands for the downstream cylinder.

\begin{figure}[h!]
 \centering
 \includegraphics[scale=.8]{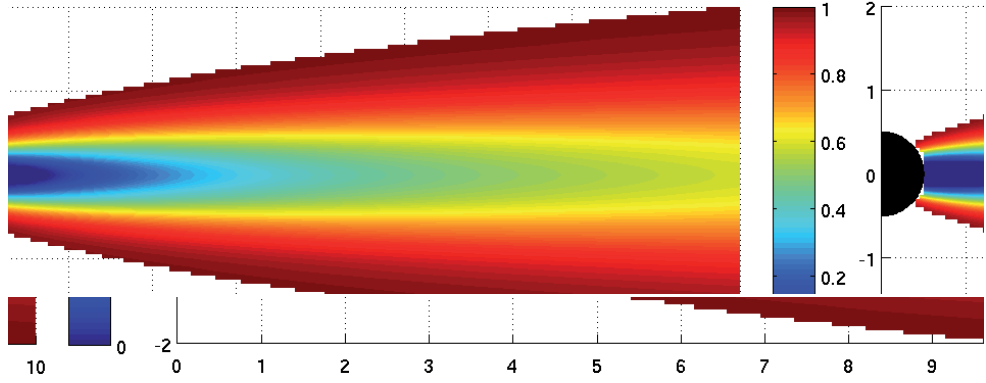}
 \caption{\it Velocity deficit behind a cylinder, according to \cite{Blevins2004}}
 \label{wake}
\end{figure}

\bigskip
For considering these wake effects, \cite{Blevins2004} proposes to use the formula of the velocity deficit, first established by \cite{Schlichting1987}, and adjusted to experiments (figure \ref{wake}):

\begin{equation} \label{wake_eq}
 u(x,y)=U_\infty \left[ 1- a_1 \sqrt{\frac{C_D \; D}{x}} \exp \left( a_2 \frac{y^2}{C_D \; D \; x} \right) \right]
\end{equation} 

with $a_1=1.2$ and $a_2=13$.

According to \cite{Blevins2004}, the equation \ref{wake_eq} is valid when $x$ is more than a few cylinder diameters. The hypothesis is that the drag force on the downstream cylinder is reduced from that in the free stream because the velocity in the wake is lower than the free stream velocity. So, it seems to be reasonable to estimate this force on a cylinder in a wake with the local wake velocity:

\begin{equation}
 C_{D_2}(x,y) = C_{D_1} \left( \frac{u(x,y)}{U_\infty} \right)^2 = C_{D_1} \left[ 1- a_1 \sqrt{\frac{C_{D_1} D}{x}} \exp \left( - a_2 \frac{y^2}{C_{D_1} D \; x} \right) \right]^2
\end{equation}

with, according to \cite{Blevins2004}, $a_1=1$ and $a_2=4.5$.

\bigskip
By admitting Price's hypothesis \cite{price1976}, which postulate that the lift is proportional to transverse gradient of drag $C_{L_2} \propto \displaystyle \frac{dC_{D_2}}{d(y/D)}$, the lift coefficient could be given:

\begin{equation}
 C_{L_2}(x,y) = a_3 \frac{C_{D_1}}{C_{D_2}(x,y)} \frac{x}{y} \left[ 1- a_1 \sqrt{\frac{C_{D_1} D}{x}} \exp \left( - a_2 \frac{y^2}{C_{D_1} D \; x} \right) \right] \sqrt{\frac{C_{D_1} D}{x}} \exp \left( - a_2 \frac{y^2}{C_{D_1} D \; x} \right)
\end{equation}

with, according to \cite{Blevins2004}, $a_1=1$, $a_2=4.5$, as before, and $a_3 = -10.6$.

\bigskip
An implicit hypothesis is made here: these formulas, established for fixed cylinders, are supposed to be true for each time step. So the drag and lift on the downstream cylinder have to be calculated for each time step, especially because of the time-dependent distances $x=x_2-x_1$ and $y=y_2-y_1$ between the two cylinders.

In addition, parameters like $C_{m_2}$, $\omega_{st_2}$, $C_{Lf_2}$ and $C_{Db_2}$ have to be calculated with the local velocity from equation \ref{wake_eq}. For the stability of the code, this velocity is considered constant and calculated from the initial position of the upstream cylinder.

Moreover, experimental data has shown that from a certain velocity, the motion of the downstream cylinder is chaotic and the oscillation frequency which is used in the calculus of the $C_{Db_2}$ is not really defined. In fact, the transverse motion of this cylinder is not periodic enough for generating an additional drag. This phenomenon is directly linked by the turbulence on the wake of the upstream cylinder. So, with these considerations, the blockage drag of the downstream cylinder is chosen equal to zero.

%%%%%%%%%%%%%%%%%%%%%%%%%%%%%%%%%%%%%%%%%%
\subsection{Link with experimental data}
%%%%%%%%%%%%%%%%%%%%%%%%%%%%%%%%%%%%%%%%%%

To compare this model with our experimental data \cite{Gaurier2007b}, we have to link it to a pendulum motion of a rigid cylinder elastically mounted in a flow. Considering little angles, we can transform the pendulum equations in linear translation equations. To do this, we have to use, in the linear equation, the mass and the stiffness linked with the slice of the cylinder considered and located at $z=1.348 \; m$. We use:

\begin{equation}
 m=\frac{I}{z^2} \quad \text{and} \quad k = \frac{K}{z^2}
\end{equation} 

\noindent with $I$ the moment of inertia, $K$ the angular stiffness and $m$ the equivalent mass and $k$ the equivalent stiffness.We use a structural damping coefficient $\zeta = 1 \%$ in agreement with experimental free decay test in calm water.

%%%%%%%%%%%%%%%%%%%%%%%%%%%%%%%%%%%%%%%%%%
\subsection{Algorithm}
%%%%%%%%%%%%%%%%%%%%%%%%%%%%%%%%%%%%%%%%%%

The numerical scheme used to solve this problem is the implicit numerical differentiation formulas \texttt{ode15s} of orders 1 to 5, from Matlab and especially designed for stiff systems \cite{Ashino2000}.

To solve equations \ref{struct_oscil} and \ref{fluide_oscil}, we need also to know $C_{Db}$ and $C_{m}$, which are dependant on the amplitude $A$ and of the frequency of transverse oscillation $f_{ex}$. To solve this dependency problem, the algorithm used makes iterative loops while $C_{Db}(t+dt) - C_{Db}(t)$ and $C_{m}(t+dt) - C_{m}(t)$ are greater than a certain value, fixed here at 0.01. Then, when $C_{Db}$ and $C_m$ are converged, the stream velocity is incremented and the motion achieved.

%%%%%%%%%%%%%%%%%%%%%%%%%%%%%%%%%%%%%%%%%%
\section{Comparison with experiments}
%%%%%%%%%%%%%%%%%%%%%%%%%%%%%%%%%%%%%%%%%%

First, a free decay test is done in order to check the natural frequency calculated and to give a initial validation of the model for one cylinder \cite{Cebron2008}. The result is not presented here, but shows an agreement for both frequency and amplitude between the model and the experimental results.

The mean hydrodynamical coefficients calculated by the model during this test are: $C_m = 0.98$, $S_t = 0.215$ and $C_{Db} = 0$. These values correspond to the classical bibliographical results.

%%%%%%%%%%%%%%%%%%%%%%%%%%%%%
\subsection{Free cylinder in a flow}
%%%%%%%%%%%%%%%%%%%%%%%%%%%%%

To validate the model in a large range of reduced velocity, the complementary characteristics: mean and standard deviation displacements can be compared between experiments and model results.

\begin{figure}[h!]
\centering
\subfigure[\textit{mean}]{\label{mean1}
\includegraphics[scale=.45]{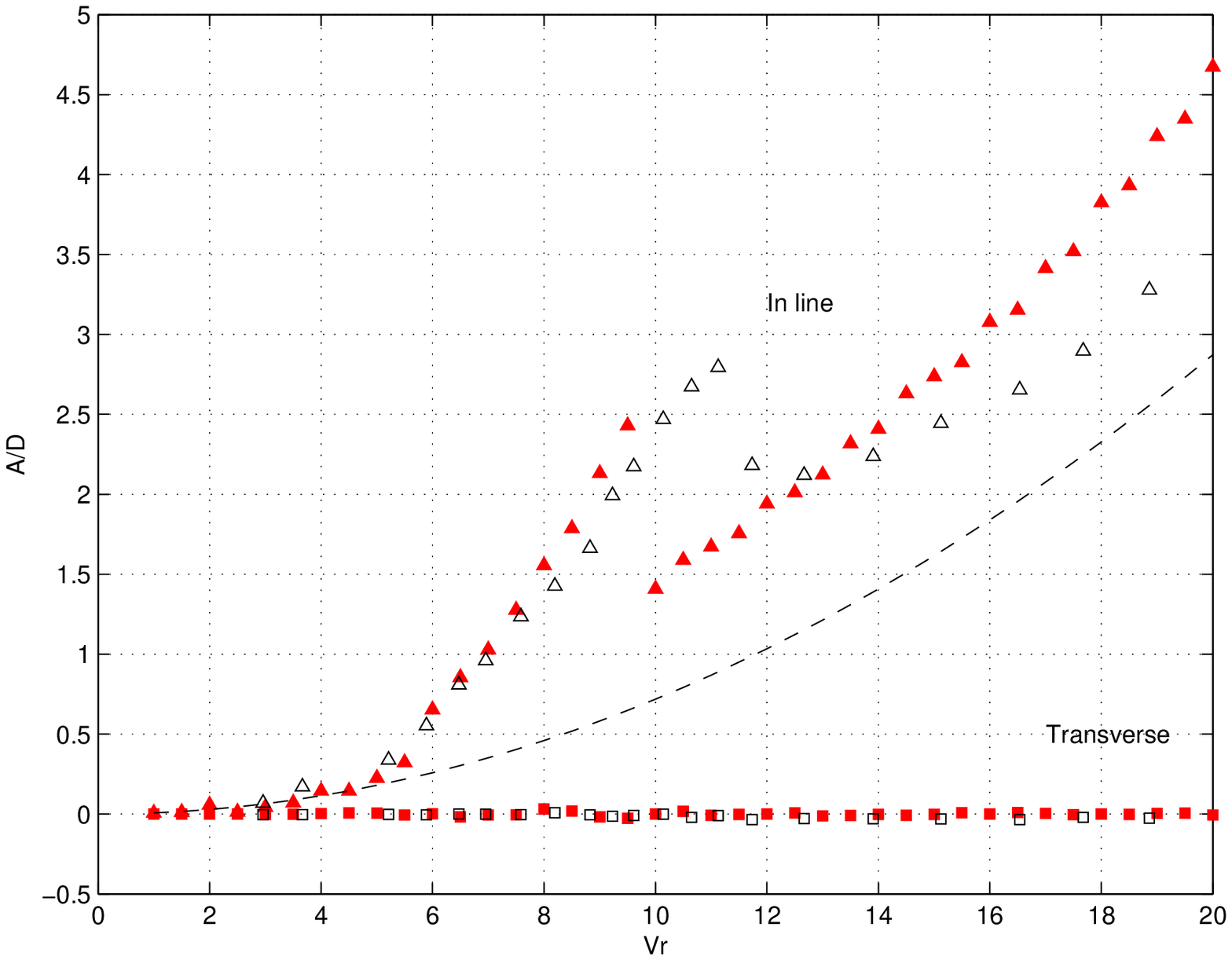}}
\subfigure[\textit{standard deviation}]{\label{std1}
\includegraphics[scale=.45]{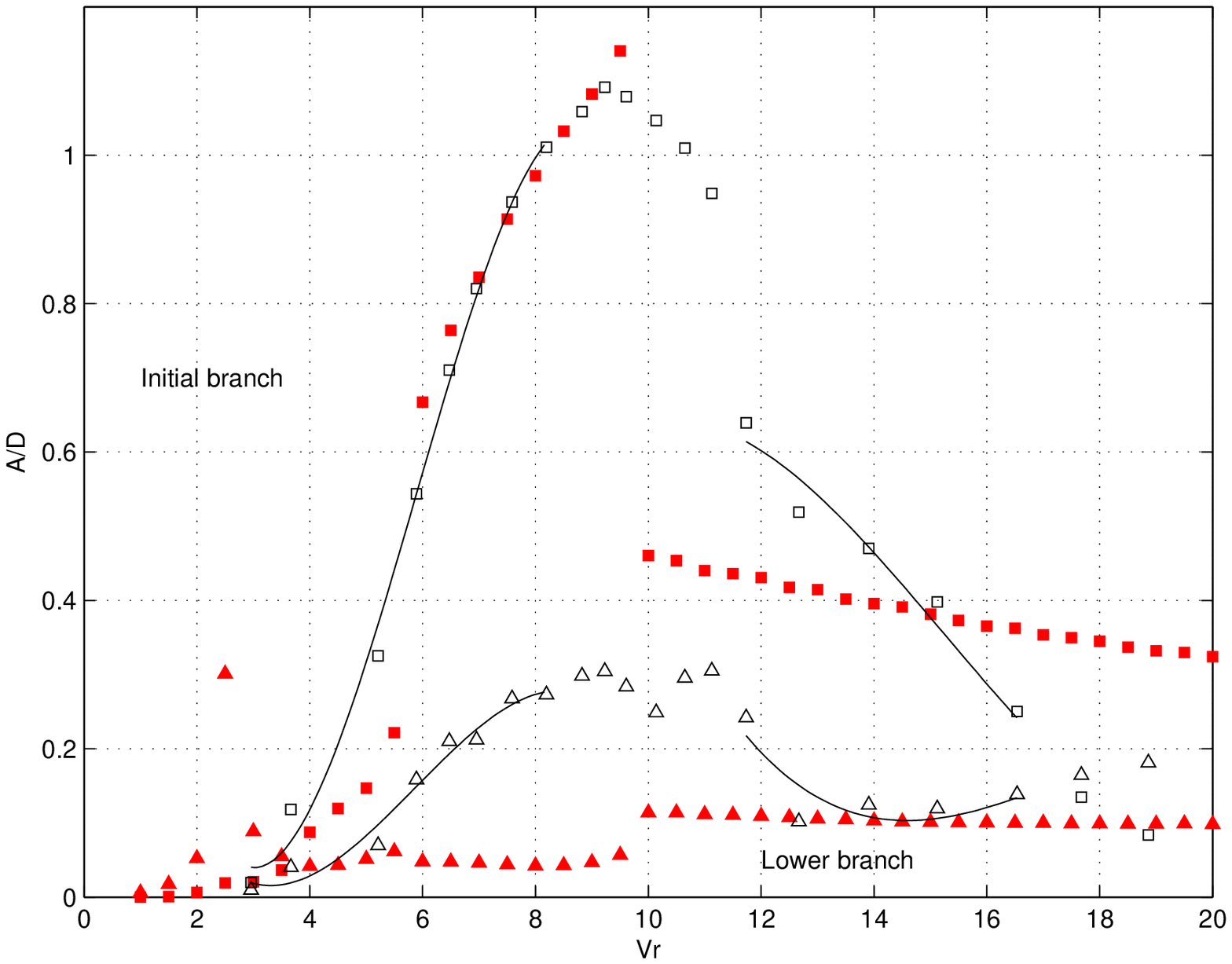}}

\caption{\it Displacements of the upstream cylinder. Transverse oscillations: $\blacktriangle$ model, $\vartriangle$ experiments. In-line oscillations: $\blacksquare$ model, $\square$ experiments. The dash line is the quasi-static result.}
\label{C1}
\end{figure}

On figure \ref{mean1}, the mean transverse displacement of the upstream cylinder is of course null for all the reduced velocities, contrary to the mean in-line displacement, which is always increasing with the velocities. The change on slope at $V_r = 10$ comes from the end of the lock-in. This sudden gap appears at a lower reduced velocity for the model (at $V_r = 10$) than for experiments (at $V_r = 11$).

On figure \ref{std1}, the three branches introduced by \cite{Jauvtis2003} are plotted on the transverse displacements: the \textit{initial branch} for $Vr < 6$, the \textit{upper branch} for $6 \leq V_r < 10$ and the \textit{lower branch} for $V_r \geq 10$. Comparing with the experimental data, the gap between the \textit{upper branch} and the \textit{lower branch} appears at $V_r = 10$ for the model, whereas it appears at $V_r = 11$ experimentally. This difference comes from the instabilities observed during experiments for these velocities: the transition from the upper to lower branch involves an intermittent switching between two modes \cite{Gaurier2007b}.

Finally, in-line r.m.s. displacements seem to be less in agreement with the experimental data than the previous data. However, the comparison with bibliographical results like \cite{Jauvtis2004} shows that the model reproduce correctly the in-line r.m.s. displacements for $V_r < 4$, including the amplification of the motion at $1.7<V_r<2.3$ named the \textit{second instability region} by \cite{Sarpkaya2004}. Model results are also in relative agreement for $V_r > 10$. However, these displacements are not reproduced during the lock-in. The origin of this difference comes from the amplification of the amplitude of $C_{Df_{01}}$ which is not take into account here. Indeed, contrary to the amplitude of the fluctuating lift coefficient $C_{Lf_{01}}$, no bibliographical data was found on this subject. So, this coefficient is considered constant in the model.

\begin{figure}[h!]
\centering
\subfigure[\textit{mean}]{\label{mean2}
\includegraphics[scale=.45]{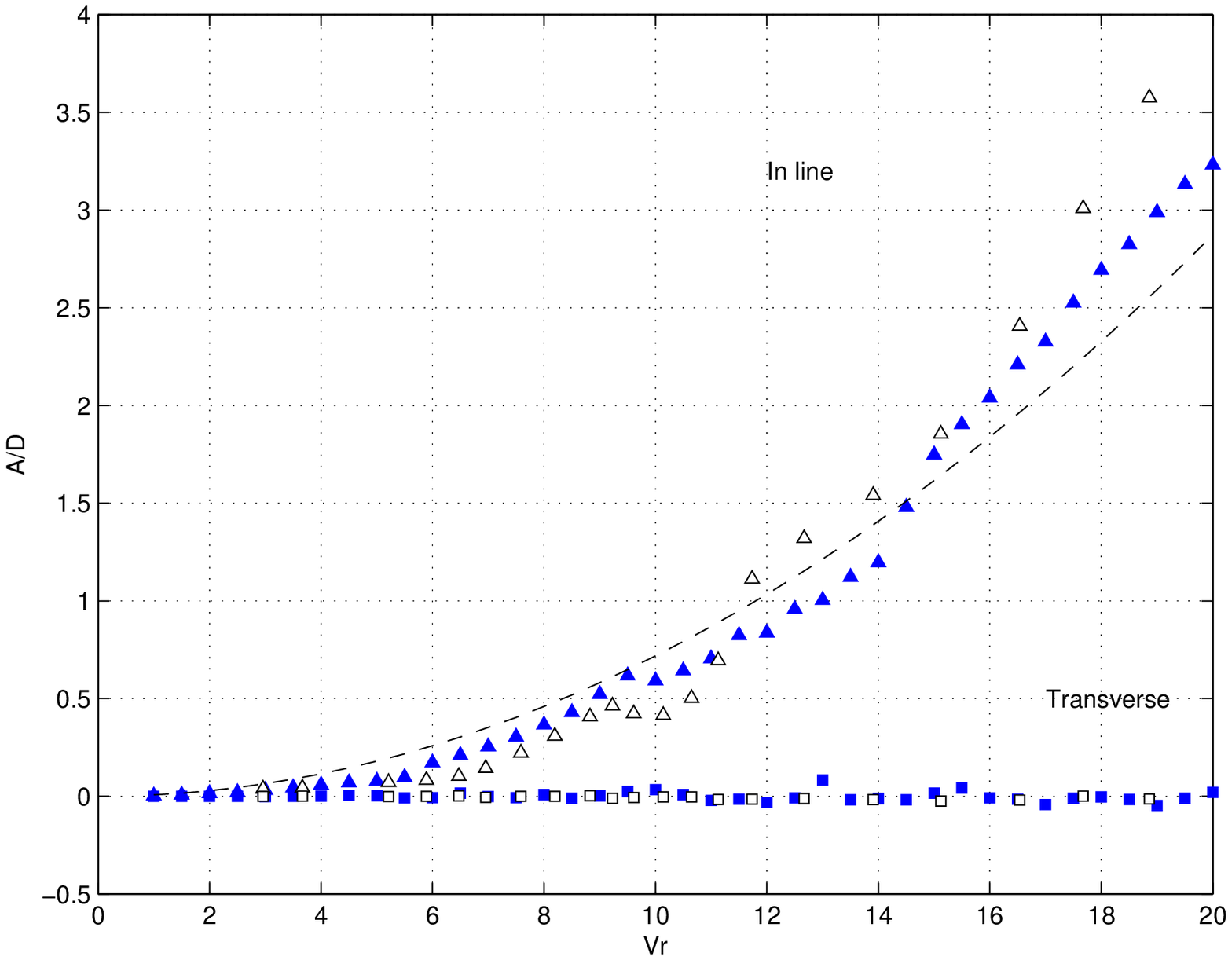}}
\subfigure[\textit{standard deviation}]{\label{std2}
\includegraphics[scale=.45]{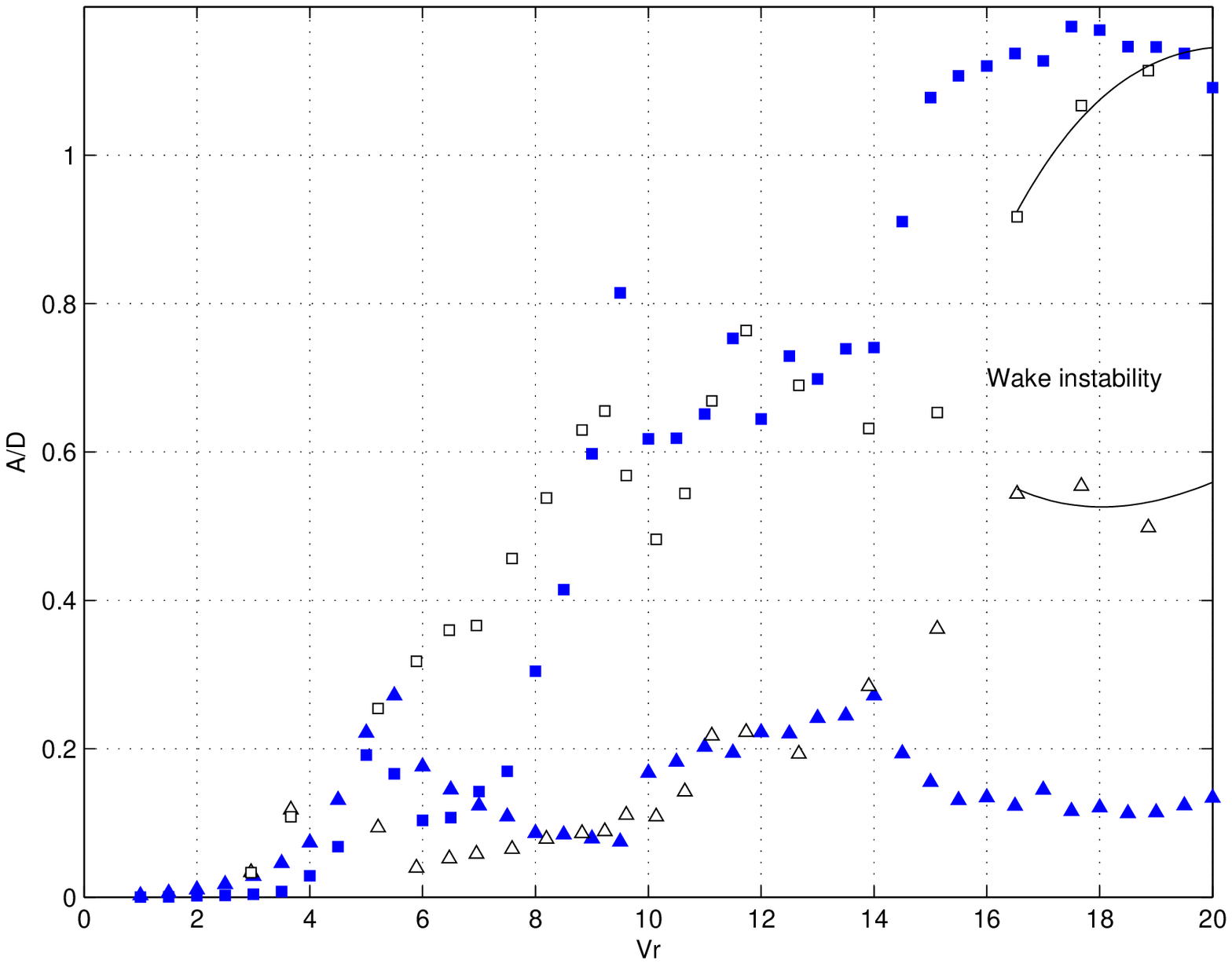}}

\caption{\it Displacements of the downstream cylinder. Transverse oscillations: $\blacktriangle$ model, $\vartriangle$ experiments. In-line oscillations: $\blacksquare$ model, $\square$ experiments. The dash line is the quasi-static result.}
\label{C2}
\end{figure}

\bigskip
For the case presented here, the downstream cylinder is initially placed at $5D$ in line behind the upstream cylinder. On figure \ref{mean2}, the mean transverse displacement of the downstream cylinder is also always null because the two cylinders are initially in tandem arrangement. The mean in-line displacement is increasing monotonically with the velocity but shows a short plate around $V_r = 10$ when the upstream cylinder's lock-in stops. As explained in paragraph \ref{hydro}, there is no amplification of the mean drag, and so of the mean in-line displacement: the slope corresponds to the quasi-static curve. The small differences observed for $12<V_r<14$ and $V_r>18$ come probably from the differences noticed in the in-line r.m.s. displacements of the upstream cylinder.

On the figure \ref{std2}, the numerical results are quite different from the experimental ones. The standard deviation of the displacement is in agreement especially for $8<V_r<14$. For the highest velocities, the wake instabilities are rather well represented by the transverse motions, but not for the in-line displacements. This difference comes essentially by the lack of data on the fluctuating amplitude of the drag force $C_{Df_{01}}$ and $C_{Df_{02}}$ for both cylinders. It could be very interesting to make new experimental trials to measure these forces on two fixed and rigid cylinders, with many initial positions, in order to determine precisely the evolution of these forces in function of the relative position of the cylinders.

%%%%%%%%%%%%%%%%%%%%%%%%%%%%%%%%%%%%%%%%%%
\section{Conclusion}

The behaviour of two cylinders in tandem arrangement at $5D$ and submit to a current had been modeled by a phenomenological model based on Van Der Pol oscillators. After the presentation of the model and the description of the used parameters, we have compared the model results with some experimental data. The comparison is relatively in agreement for the mean and standard deviation of the two dimensional cylinders motions. The differences come essentially from the poor set of data for the fluctuating drag coefficient. Some specific experimental tests should be conducted to solve this problem.

Despite those imperfections, other initial configurations will be tested and compared between experimental data and this simple model, to extend and determine the limitations of this kind of two dimensional code.

%%%%%%%%%%%%%%%%%%%%%%%%%%%%%%%%%%%%%%%%%%
% \section{Conclusion}
\bibliography{riserbib}
\bibliographystyle{acm}

\end{document}